\documentclass[fleqn,usenatbib]{mnras}

\usepackage{newtxtext,newtxmath}
\usepackage{xspace}

\usepackage{graphicx}	
\usepackage{amsmath}	
\usepackage{float}

\title{On the Orbital Velocity of Isolated Galaxy Pairs: II Accurate MOND Predictions. }
\author[R. Scarpa et al.]{Riccardo Scarpa,$^{1,2}$\thanks{E-mail:
riccardo.scarpa@gtc.iac.es}
Renato Falomo,$^{3}$
Aldo Treves$^{4,5}$
\\
$^1$ Instituto de Astrofisica de Canarias, C/O Via Lactea, s/n, E-38205 La Laguna, Tenerife, Spain\\
$^2$ Departamento de Astrofsica, Universidad de La Laguna, E-38206 La Laguna, Tenerife, Spain\\
$^3$INAF - Osservatorio Astronomico di Padova, vicolo dell’Osservatorio 5, I-35122 Padova, Italy\\
$^4$ Universit\`a dell’Insubria, via Valleggio 11, I-22100 Como, Italy\\
$^5$ INAF - Osservatorio Astronomico di Brera, via Bianchi 46, I-23807 Merate (Lecco), Italy\\
}

\date{Last Rev. 17 January  2022}

\begin{document}
\label{firstpage}
\maketitle

\begin{abstract}

Examining a catalogue of isolated galaxy pairs, a preferred orbital
intervelocity of $\sim$ 150 km/s was recently reported. This discovery
is difficult to reconcile with the expectations from Newtonian
numerical simulations of cosmological structure formations. In a
previous paper we have shown that a preferred intervelocity for galaxy
pairs is expected in Modified Newtonian Dynamics (MOND). Here a
detailed analysis of the MOND predictions is presented, showing that a
remarkable agreement with the observations can be achieved for a mass
to light ratio M/L$\sim$1 in solar units. This agrees with the
expectations for a typical stellar population, without requiring
non-baryonic dark matter for these systems.
\end{abstract}


\section{Introduction}

Galaxy pairs represent an important probe to investigate the behaviour of gravity under particular conditions that cannot be found in the laboratory or in the very local Universe. For typical masses of $\sim$ 10$^{11}$ $M_\odot$ and separation of $\gtrsim 100$ kpc the acceleration of gravity between the two galaxies is indeed below $10^{-10}$ cm s$^{-2}$, a regime difficult to probe in the solar system or other stellar structures. 

\begin{figure}
\centering
\includegraphics[scale=0.53]{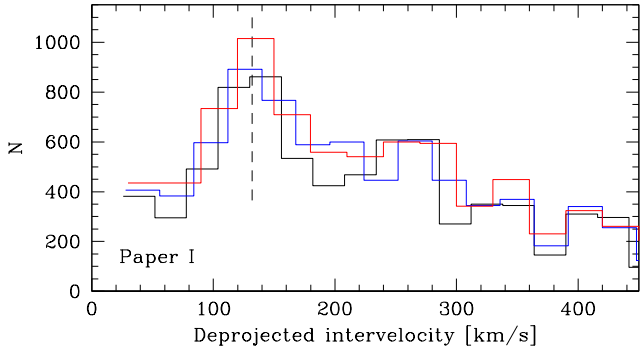}
\caption{Distribution of deprojected 3D intervelocities for the 8571 pairs with isolation parameter $\rho>5$ (see paper I) extracted from an extended version of isolated galaxy pairs catalog, adapted from Paper I. Three different binning were used to show the peak  at $\sim 150$ km/s is clearly present and stable. The vertical dashed line mark the peak center at 132 km/s (see text).}
\label{peakpos}
\end{figure}

\cite{NC18a} constructed a catalog of $\sim$ 13000 close-by ($0.01<z<0.05$) isolated galaxy pairs (IGP), extracted from the HyperLeda extragalactic database \citep{Makarov14}. Starting from the observed radial velocity  difference (intervelocity hereafter) of the components of the pair, and using a technique of statistical deprojection described in \cite{NC18b}, they were able to statistically reconstruct the 3D velocity distribution. The remarkable result was found, that a region of preferred intervelocities with a peak at $\sim$ 150 km/s does exist \citep{NC20}. Considering a larger version of the IGP catalog, containing $\sim$ 16500 pairs, extracted from a more recent version of HyperLeda, \cite{SFT22}, hereafter Paper I, confirmed the presence and position of the peak (see Fig.\ref{peakpos}). 

Current Newtonian simulations of galaxy structure formation do not seem to predict a narrow range of intervelocities for galaxy pairs (e.g. \cite{Moreno13} and references therein). However, this is a recently discovered feature and dedicated numerical simulations are required to fully address the issue.

Meanwhile, in paper I it was shown that a narrow region of orbital velocities of galaxy pairs is  expected within the framework of modified Newtonian dynamics (MOND, \cite{Milgrom83a,Milgrom83b,Milgrom83c}). This 
because the orbital velocity V does not depend on the separation of the galaxies within the pair, and is only mildly linked to the mass M of the galaxies ($V\propto M^{1/4}$, see below).

Here we revisit these findings and present a detailed discussion of the MOND predictions. We focus on the treatment of the two body problem, moving from the initial approximate formula proposed in \citet{Milgrom83c}, to the  rigorous formulation discussed in \citet{Milgrom94}. The peak position of the 3D intervelocity distribution is then used to constrain the mass-to-light M/L ratio of the population under examination.

\section{The Peak of intervelocity distribution}

To better characterize the properties of the inter-velocity peak we performed a detailed analysis of its position and width. Of the various  deprojection techniques discussed by  \citep{NC18b}, here and in paper I the "differences on adjacent constant bins" algorithm is used for being the one with the simplest physical interpretation. This technique, and all the others for that matter, requires a monotonically decreasing distribution as input.  
This implies the deprojection can be performed only adopting sufficiently large bins so that the statistical fluctuations in the observed intervelocity distribution are smoothed out.

\begin{figure}
\centering
\includegraphics[scale=0.7]{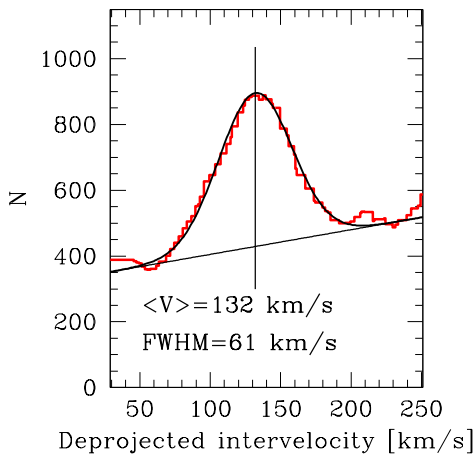}
\caption{ Gaussian fit (thick black line) of the region around the peak as seen in the re-sampled 3D deprojected intervelocity distribution (red histogram). The inclined line represent the adopted background level due to unbound pairs (see also Paper I). The vertical line marks the center of the peak.
}
\label{peakfit}
\end{figure}

As a consequence, in the resulting 3D deprojected velocity distribution the peak position depends somewhat  on the chosen  bin size (see Figure \ref{peakpos}), and the true peak position is loosely defined. We therefore performed the deprojection for ten different values of bin size from 23 to 32 km/s with a step of 1 km/s. The resulting 3D velocity distributions were then re-sampled with a common step of 1 km/s and averaged together (Fig. \ref{peakfit}). A  Gaussian fit of the region around the peak gives a best fit position of $132\pm 5$ km/s and FWHM $61\pm 5$ km/s.

\section{The two body problem in MOND}

Flat rotation curves of galaxies indicate that the mass distribution within galaxies is such that these structures do enter in MOND regime at distances from their center smaller than their size. Because of this, the interaction between pair of galaxies always occur in the so called deep MOND regime, in which all relevant accelerations are well below the MOND constant $a_0 = 10^{-8}$ cm s$^{-2}$. This is important because it removes the uncertainties on the unknown interpolation function needed in the Newtonian-MOND intermediate regime, and because in the deep MOND limit the two body problem can be solved exactly \citep{Milgrom94}, making galaxy pairs particularly interesting for the study of dynamics in the low acceleration limit. 

\begin{figure}
\centering
\includegraphics[scale=0.7]{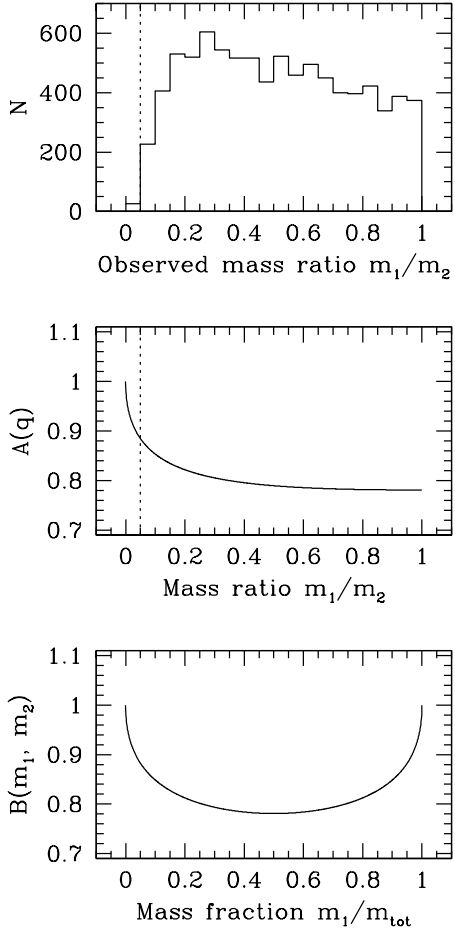}
\caption{ {\bf Upper panel:} The mass ratio for pairs from the selected galaxies from Paper I. The vast majority of pairs do have mass ratio above 0.05 (dashed line). This line is reported also in the following panel to mark the region of the functions A relevant to our samples.
{\bf Middle panel:} The function A(q) over its full range of values. Note that q has to be calculated so that the mass ratio is always $\leq 1$ to ensure the force between the two masses is the same. {\bf Bottom panel:} The function B($m_1,m_2)$. The minimum corresponds to the case of equal masses.}
\label{functions}
\end{figure}

In deep MOND regime the expression for the two body force becomes \citep{Milgrom94}: 

\begin{equation}
F(m_1,m_2,s)=\frac{m_1 m_2}{s}\sqrt{\frac{Ga_0}{m_1+m_2}} A(\frac{m_1}{m_2})
\end{equation}

with 
\begin{equation}
A(q) = \frac{2\sqrt{1+q}}{3q}[(1+q)^\frac{3}{2}-q^\frac{3}{2}-1]
\end{equation}

where F is the gravitational force between two masses $m_1$ and $m_2$, $s$ their separation, $q$ is the ratio of the smaller mass to the larger one ($q\leq 1$), and G the gravitational constant. 
The value of the coefficient A(q), which is symmetric with respect to the two masses, varies little in its full range, from 0.781 at $q=1$, to 1 for $q=0$ (Fig. \ref{functions}). 
It was later shown that the above expression is valid for all different flavours of modified dynamics \citep{Milgrom14}. 

In the simple case of two equal masses $m_1=m_2=m_{tot}/2$ on circular orbits of radius r=s/2, then $V_1= V_2=V$ and the intervelocity $\Delta V = 2V$ reads:

\begin{equation}
\Delta V^4 = 0.610 G ~a_0 ~m_{tot}.
\end{equation}

The numerical factor in this equation, which is exact in deep MOND regime, is a factor $\sim 2$ smaller than the one appearing in the approximate formula presented by \cite{Milgrom83c}, formula that was used in our paper I.

In the case of circular orbits, eq. 1 and 2 allow to derive a general expression for the intervelocity for two bodies of different masses. The relation becomes:

\begin{equation}
\Delta V^4 = G ~a_0 ~m_{tot} ~B^2(m_1,m_2) 
\end{equation}

with

\begin{equation}
B(m_1, m_2) =  \frac{2}{3}\left(\frac{1-x_1^{3/2}-x_2^{3/2}}{x_1~x_2}\right)
\end{equation}

where $x_i=m_i/m_{tot}$ is the mass fraction of each body, and again the separation no longer appears. In case of non circular orbits, the instantaneous velocity  $\Delta V$ has to be replaced by the average velocity and the relation become $<\Delta V^2>^2 = G a_0 m_{tot} B^2$. So non-circular orbits would lead to a broadening of the distribution of instantaneous $\Delta V$, but would not shift the position of $\Delta V$ \citep{Milgrom14}.

Similarly to A(q), the function B($m_1,m_2$) varies little within its full range (Fig. \ref{functions}). The shape of B is such that the intervelocity is minimal when masses are equal. The maximum B=1 corresponds to the case in which one mass is negligible, as in the case of a star orbiting a galaxy, and consistently the usual expression employed to study galaxy rotation curves is recovered.

Based on the above formulae it is therefore possible to predict the expected 3D intervelocity distribution for a sample of bound galaxy pairs, and compare it to the observed data.

\section{ MOND expectation for the 3D intervelocity distribution }

\begin{figure}
\hspace{-0.5cm}
\includegraphics[scale=0.5]{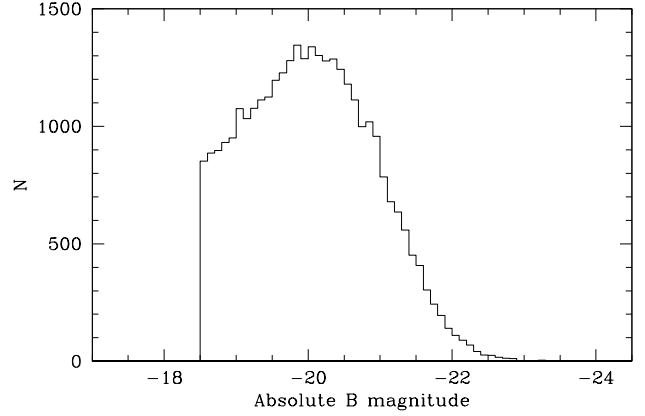}
\caption{Absolute B-band magnitude distribution for all selected galaxies.  Note the sharp cut off at M$= -18.5$ applied to the catalog.}
\label{fig_absoluteMags}
\end{figure}

\begin{figure}
\hspace{0.5cm}
\includegraphics[scale=0.8]{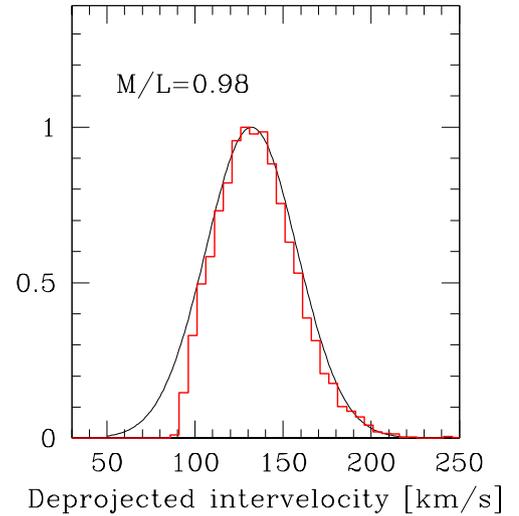}
\caption{Deep MOND intervelocity (histogram) compared to the observed peak position (Gaussian) as from Fig. \ref{peakfit}. The best fit M/L ratio is given.}
\label{fig_deep_mond}
\end{figure}

The observed luminosity of the pairs of galaxies  (Fig. \ref{fig_absoluteMags}) can be converted into mass adopting a fixed M/L ratio for the whole population. As it is not possible to know which pairs are bound and belong to the peak and which do not, it is implicitly assumed that the M/L does not varies significantly for bound and unbound galaxy pairs. Equations 4 and 5 are then used, appropriately considering the mass fraction of the components (Fig. \ref{functions}).  The result is then directly compared to the deprojected intervelocity distribution.
We allowed M/L to vary so to match the observed distribution .
It is found that for reasonable values of M/L the deep MOND prediction matches both the position of the peak and its overall shape (Fig. \ref{peakfit}). The agreement is good  (Fig. \ref{fig_deep_mond}), and the best fit M/L$\sim 1$ is in line with the expectation for a typical stellar population. 

We remark that  the very same analysis of the \cite{NC20} sample, limited as in our case to pairs with isolation parameter $\rho>5$ (see paper I), results in an intervelocity peak centered at $141\pm 5$ km/s (FWHM$\sim 50$ km/s) and a best fit M/L of $\sim 1.35$,  somewhat larger than what is found for our sample. This gives an indication of the uncertainties associated to the parameters derived with this method.

\section{Conclusions}

We presented a strict consistency of the position and width of the peak of 3D intervelocities of isolated galaxy pairs with the deep MOND predictions, when M/L$\sim 1$ is assumed, and therefore no dark matter is required in these systems. This consistency reinforces the proposal that MOND is the key for interpreting the peak, rather than hypothesize a complex process of structure formation within Newtonian dynamics that force galaxies pairs to preferentially end up with a well defined orbital velocity.

We recall, however, that these results are based on the IGP catalog defined by \cite{NC18a} and its extension 
using the updated HyperLeda database as presented in paper I.
The two samples largely overlap, therefore to further confirm the presence and the properties of the intervelocity peak an independent data set of galaxy pairs is required. This  would imply to extend the study to higher redshifts and fainter  magnitude of galaxies to ensure a good level of homogeneity in the sample (see also discussion in \cite{NC20}).
Another possibility would be to focus on galaxies of specific Hubble types, studying separately spirals from ellipticals. These galaxies are thought to have gone through different dynamical evolutionary paths. However, masses being equal, within the MOND framework, spirals and ellipticals pairs should exhibit the same preferred intervelocities. 
Finally, if a large sample of pair of quasars could be build, their massive host galaxy would extend the present result to even larger masses, also ensuring the perturbing effect of other nearby galaxies on the orbital velocity would be less important.

\vspace{1cm}

{\bf Data Availability:} The data supporting the findings of this study are openly available at HyperLeda web site (http://leda.univ-lyon1.fr).

\section*{Acknowledgments}

We are very grateful to Moti Milgrom for important and constructive suggestions on the manuscript. We also thank the anonymous referee for many useful comments on the original version of the manuscript.


\begin{thebibliography}{}

\bibitem[\protect\citeauthoryear{Makarov et al.}{2014}]{Makarov14}Makarov D., Prugniel P., Terekhova N., Courtois, H. \& Vauglin I. 2014, A\&A, 570, 13
\bibitem[\protect\citeauthoryear{Nottale and Chamaraux}{2018a}]{NC18a} Nottale L. and Chamaraux P. 2018a, Astrophysical Bulletin, 73, 310
\bibitem[\protect\citeauthoryear{Nottale and Chamaraux}{2018b}]{NC18b} Nottale L. and Chamaraux P. 2018b, A\&A, 614, 45 
\bibitem[\protect\citeauthoryear{Nottale and Chamaraux}{2020}]{NC20}Nottale L. and Chamaraux P. 2020, A\&A 641, 115
\bibitem[\protect\citeauthoryear{Milgrom }{1983a}]{Milgrom83a} Milgrom M. 1983a, ApJ 270, 365 
\bibitem[\protect\citeauthoryear{Milgrom }{1983b}]{Milgrom83b}------. 1983b, ApJ 270, 371
\bibitem[\protect\citeauthoryear{Milgrom }{1983c}]{Milgrom83c}------. 1983c, ApJ 270, 384
\bibitem[\protect\citeauthoryear{Milgrom }{1994}]{Milgrom94}------. 1994, ApJ 429, 540
\bibitem[\protect\citeauthoryear{Milgrom }{2014}]{Milgrom14}------. 2014, PhRvD, 89, 4016 
\bibitem[\protect\citeauthoryear{Moreno et al. }{2013}]{Moreno13}Moreno J., Bluck A. F. L., Ellison S. L., et al. 2013 MNRAS 436, 1765
\bibitem[\protect\citeauthoryear{Scarpa  Falomo and Treves}{2022}]{SFT22} Scarpa R., Falomo R., and Treves A. 2022, MNRAS 510 2167 (Paper I)
\end{thebibliography}
\end{document}